\documentclass[aps,pra,twocolumn,superscriptaddress]{revtex4-1}
\usepackage{graphicx}
\usepackage{color}
\usepackage{amssymb}
\begin{document}
\title{A simple anisotropic three-dimensional quantum spin liquid with fracton topological order}
\author{O. Petrova}
\affiliation{
D\'epartement de physique de l'ENS, \'Ecole normale sup\'erieure / PSL Research University, CNRS, 24 rue Lhomond, 75005 Paris, France}
\author{N. Regnault}
\affiliation{
Laboratoire Pierre Aigrain, D\'epartement de physique de l'ENS, \'Ecole normale sup\'erieure, PSL Research University, Universit\'e Paris Diderot, Sorbonne Paris Cit\'e, Sorbonne  Universit\'es, UPMC Univ. Paris 06, CNRS, 75005 Paris, France
}

\begin{abstract}
We present a three-dimensional cubic lattice spin model, anisotropic in the $\hat{z}$ direction, that exhibits fracton topological order. The latter is a novel type of topological order characterized by the presence of immobile pointlike excitations, named fractons, residing at the corners of an operator with two-dimensional support. As other recent fracton models, ours exhibits a subextensive ground state degeneracy: On an $L_x\times L_y\times L_z$ three-torus, it has a $2^{2L_z}$ topological degeneracy, and an additional non-topological degeneracy equal to $2^{L_xL_y-2}$. The fractons can be combined into composite excitations that move either in a straight line along the $\hat{z}$ direction, or freely in the $xy$ plane at a given height $z$. While our model draws inspiration from the toric code, we demonstrate that it cannot be adiabatically connected to a layered toric code construction. Additionally, we investigate the effects of imposing open boundary conditions on our system. We find zero energy modes on the surfaces perpendicular to either the $\hat{x}$ or $\hat{y}$ directions, and their absence on the surfaces normal to $\hat{z}$. This result can be explained using the properties of the two kinds of composite two-fracton mobile excitations.
\end{abstract}

\maketitle

\section{Introduction}

Topological order \cite{PhysRevB.40.7387} has arguably been one of the most exciting developments in physics in the last couple of decades. First, it is fascinating from the fundamental science point of view, describing types of order beyond the conventional theory of spontaneous symmetry breaking. In place of broken symmetries, phases are identified by more exotic properties such as ground state degeneracies that depend on the system's topology, and anyonic excitations - quasi-particles with fractionalized exchange statistics \cite{PhysRevLett.49.957}. From a more applied perspective, anyons have been proposed as building blocks for the enticing prospect of topological quantum computing \cite{toricKitaev}. Such particles have originally been proposed to exist in two dimensions only, since group representation theory requires all pointlike particles to behave either as bosons or as fermions in higher dimensions. Once the pointlike requirement is relaxed, however, the concept of anyons (and with it, that of topological order) can be extended to the familiar three dimensions as well as beyond, to more exotic realms \cite{PhysRevB.78.155120,PhysRevB.72.035307}.

Many topologically ordered models that have been proposed so far can be cast into gauge theory forms, one such example being the famous toric code \cite{toricKitaev}, equivalent to a $\mathbb{Z}_2$ gauge theory. Recently, a new type of topological order has been proposed: it does not admit a gauge theory description, and is instead characterized by the presence of so-called \emph{fractons}, pointlike fractionalized quasi-particles that are strictly immobile \cite{PhysRevLett.94.040402,BRAVYI2011839,HaahCode,Yoshida,PhysRevB.92.235136,PhysRevB.94.235157,PhysRevB.94.155128,fractonFromLayers,PhysRevB.95.115139,PhysRevB.95.155133,1701.00762,1703.02973,1706.07070,1707.02308}. It is precisely the latter property that allows these phases to get around the requirement for extended ($d>0$) anyons beyond two dimensions: since fractons are immobile, there is no notion of exchange. Instead, fractons can be combined into composite excitations that propagate in subspaces whose dimensionality is reduced from that of the physical system. Such fracton models exhibit subextensive topological degeneracy.

In this article, we introduce a model defined for spins $1/2$ living on the sites of a $L_x\times L_y\times L_z$ cubic lattice. It has immobile pointlike excitations, dubbed fractons, that are located at the corners of an operator with two dimensional support. In addition to the subextensive topological ground state degeneracy that goes as $2^{2L_z}$, we obtain an additional, albeit not topologically protected, degeneracy equal to $2^{L_x\times L_y-2}$. While topological ground state degeneracies, normally defined for systems with periodic boundary conditions, are one of the hallmarks of topological order, real-life compounds normally come with a boundary. The associated edge states depend on the topological order in the bulk, making edge physics an important characteristic and a major experimental probe into the nature of a topological phase \cite{PhysRevB.41.12838}. Interestingly, imposing open boundary conditions in the $\hat{z}$ direction has no effect on the total degeneracy of the ground state, although it does change the spectrum of the excitations. Alternatively, open boundary conditions in the $\hat{x}$ ($\hat{y}$) directions result in additional exponents of $L_y\times L_z$ ($L_x\times L_z$) in the expression for the ground state degeneracy. We note the important distinction between these new zero energy surface modes, and the $2^{L_x\times L_y}$ bulk degeneracy above. In the latter case, the system can only transform from one state to another via an operator spanning its length in the $\hat{z}$ direction, whereas the surface zero energy modes are generated locally.

\section{The model}

\subsection{The cube operators}

One road to three-dimensional models lies via layering of two-dimensional ones \cite{PhysRevB.87.235122,PhysRevX.4.041043,fractonFromLayers,1701.00762,1706.07070,PhysRevB.93.205406}. In particular, different ways of coupling two-dimensional toric code \cite{toricWen,toricKitaev} layers can lead to a three-dimensional toric code (a $3+1$D $Z_2$ gauge theory) \cite{PhysRevB.72.035307,PhysRevB.78.155120}, or to a number of fracton three-dimensional topological phases \cite{fractonFromLayers,1701.00762}.

\begin{figure}[h]
\includegraphics[width=\columnwidth]{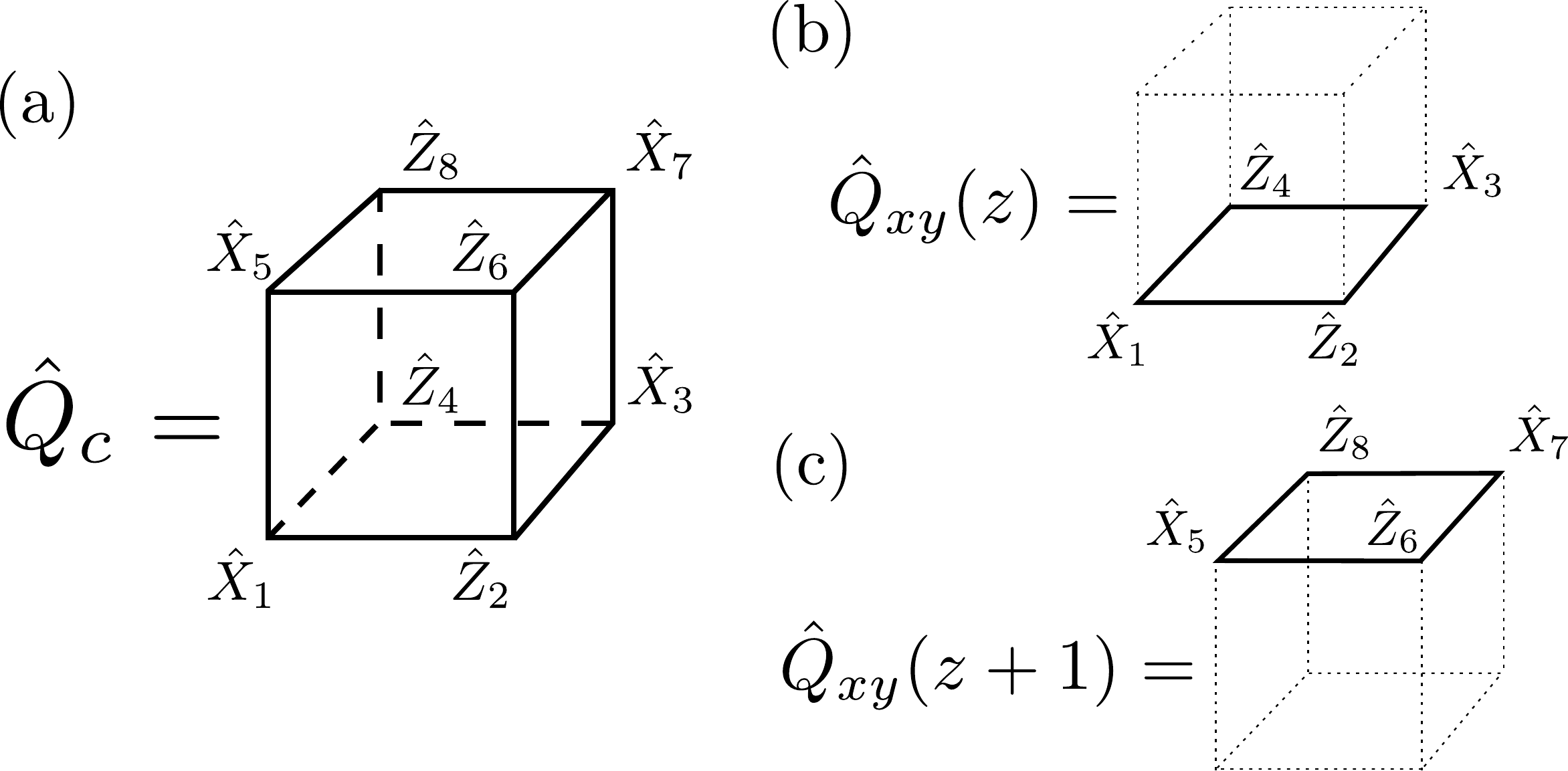}
\caption{Operator definitions for the 3D model. The cube operator (a) $\hat{Q}_c$  is the product of the two plaquette operators at adjacent $xy$ planes: (b) $\hat{Q}_{xy}(z)$ and (c) $\hat{Q}_{xy}(z+1)$.}
\label{fig:operators}
\end{figure}

The model that we present here also draws inspiration from the two-dimensional toric code, reviewed in the Appendix. We consider an $L_x\times L_y\times L_z$ cubic lattice with spins $1/2$ living on the lattice sites. For each of the $L_z$ $xy$ planes, we can define a two-dimensional toric code model. Here and below we are using capital letters $\hat{X}$, $\hat{Z}$ to denote the corresponding Pauli matrix operators. The toric code Hamiltonian involves a sum of square plaquette operators 
\begin{equation}
\hat{Q}_{xy}=\hat{X}_1 \hat{Z}_2  \hat{X}_3 \hat{Z}_4,
\label{eq:Qxy}
\end{equation} 
where the lattice sites are labeled as they appear in the definition of $\hat{Q}_{xy}(z)$, a plaquette operator located on the $z^\mathrm{th}$ $xy$ layer, shown in Fig.~\ref{fig:operators}(b).

Now consider two plaquette terms $\hat{Q}_{xy}(z)$ and $\hat{Q}_{xy}(z+1)$ centered at the same $xy$ coordinates on adjacent layers at $z$ and $z+1$ respectively as shown in Fig.~\ref{fig:operators}(b-c). Multiplying the two operators leads to an eight-spin operator 
\begin{equation}
\hat{Q}_c=\hat{Q}_{xy}(z)\times\hat{Q}_{xy}(z+1)=\hat{X}_1 \hat{Z}_2  \hat{X}_3 \hat{Z}_4 \hat{X}_5 \hat{Z}_6  \hat{X}_7 \hat{Z}_8,
\label{eq:Qc}
\end{equation} 
associated with a cube, as shown in Fig.~\ref{fig:operators}(a). All such cubic terms commute with each other. The Hamiltonian of our three dimensional model is simply the sum of all $\hat{Q}_c$ operators taken with a global minus sign:
\begin{equation}
H=-\sum_{c}\hat{Q}_c.
\label{eq:H3d}
\end{equation}
We note that the constituent $\hat{Q}_{xy}(z)$ operators defined on the $xy$ planes commute among each other as well as with all of the $\hat{Q}_c$ operators, therefore, both $\hat{Q}_{xy}(z)$ and $\hat{Q}_c$ are integrals of motion of the three-dimensional model whose Hamiltonian is given by Eq.~(\ref{eq:H3d}).

\subsection{Ground state degeneracy}

The possible eigenvalues $Q_c$ of the cubic operators $\hat{Q}_c$ are $\pm1$. In the ground state, all cubes have $Q_c=+1$. It follows that the plaquette terms $\hat{Q}_{xy}(z)$ in a given column along the $\hat{z}$ direction have their eigenvalues $Q_{xy}$ being either all $+1$ or all $-1$ (which allows us to drop the $z$ coordinate when specifying the eigenvalue of $\hat{Q}_{xy}(z)$ in a ground state). The choice of $Q_{xy}=\pm1$ per column is reflected in the extensive ground state degeneracy of the model, $\mathcal{D}$. The latter is found by calculating the difference between the total number of spins and the number of independent constraints that the system has to satisfy in its ground state. This difference is equal to $\log_2\mathcal{D}$ \cite{PhysRevA.54.1862}. In our case, the ground state constraints are that $Q_c=+1$ for all cubes. Below we treat the case of a $L_x\times L_y\times L_z$ three-torus; the open boundary conditions are discussed separately in Section \ref{sec:open}. The number of spins in our system is $L_x\times L_y\times L_z$, and so is the number of cubes, i.e.~the number of $Q_c=+1$ constraints for the ground state. However, not all of these constraints are independent. Consider a single spin flip, that is shown in Fig.~\ref{fig:3dFlip}: It flips the eigenvalues of four cube operators. There are two pairs of cubes stacked on top of one another, and these two stacks are located diagonally from each another (as in Wen's version of the toric code in two dimensions \cite{toricWen}, described in the Appendix). Because each flipped spin results in flipping the eigenvalues of $Q_c$ for two diagonal stacks, we color the columns along the $z$ direction in a checkerboard pattern, and identify the two colors of the ``board'' with two flavors of excitations, $e$ and $m$. For each flavor, excitations are created in pairs, resulting in two relations for the eigenvalues of the $\hat{Q}_c$ operators
\begin{equation}
\prod_{c\in e}Q_c=+1 \mathrm{~and~} \prod_{c\in m}Q_c=+1
\label{eq:emrelations}
\end{equation}
per each of the $L_z$ layers of cubes. Additionally, there are $L_x\times L_y-2$ relations for the columns 
\begin{equation}
\prod_{c\in \mathrm{column}}Q_c=+1,
\label{eq:crelation}
\end{equation}
whereas the values of $Q_c$ in the remaining two columns are completely determined by Eq.~(\ref{eq:emrelations}). Combining the relations Eqs. (\ref{eq:emrelations}) and (\ref{eq:crelation}) reduces the number of independent ground state constraints from the total number of cubes, resulting in an extensive ground state degeneracy $\mathcal{D}=2^{L_x\times L_y-2+2L_z}$. As will be elucidated in Section \ref{sec:excitations}, the $2^{2L_z}$ factor is topological, remnant of the toric code-like $xy$ planes. The ground states can therefore be classified by $L_z$ pairs of winding numbers, e.g. $W_{e}^{x}(z)$ and $W_{e}^{y}(z)$, in correspondence to the toric code's $W_{e}^{x}$ and $W_{e}^{y}$ defined in the Appendix.

The $\mathcal{D}=2^{L_x\times L_y-2}$ degeneracy comes from the fact that in addition to the winding numbers, the ground states are also distinguished by $Q_{xy}=\pm1$ line going through each of the $L_x\times L_y$ columns along the $\hat{z}$ direction. One way to label the eigenstate is by using eigenvalues of $\hat{Q}_{xy}$ as quantum numbers. However, there are only $L_x\times L_y-2$ independent $Q_{xy}$ values per $xy$ layer of spins. Indeed, plaquettes on each $xy$ plane are associated with two flavors, and for each flavor the eigenvalues of  $\hat{Q}_{xy}$ are flipped in pairs. Similarly to an Ising chain, in order to switch from one ground state to another, we have to flip at least $L_z$ spins. These states are locally distinguishable and therefore this degeneracy is not topologically protected. 

\subsection{Excitations}
\label{sec:excitations}

\begin{figure}
\includegraphics[width=0.5\columnwidth]{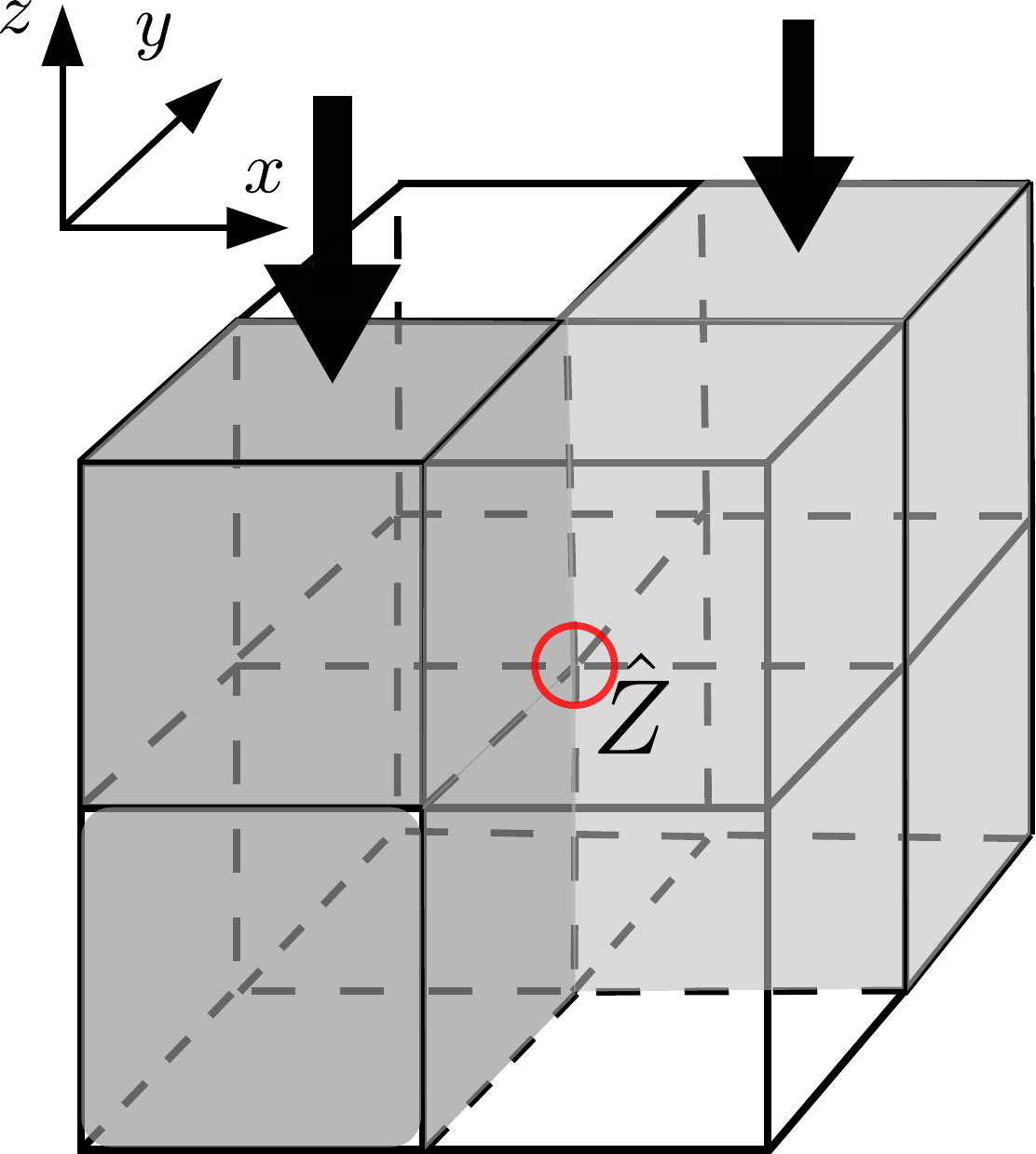}
\caption{Starting with a system in a ground state, acting on the center spin with $\hat{Z}$ creates four $Q_c=-1$ excitations: excited cubes (dark grey and light grey) are diagonal from each other in the $xy$ plane and stacked one on top of the other along the $\hat{z}$ direction.}
\label{fig:3dFlip}
\end{figure}

\begin{figure}
\includegraphics[width=\columnwidth]{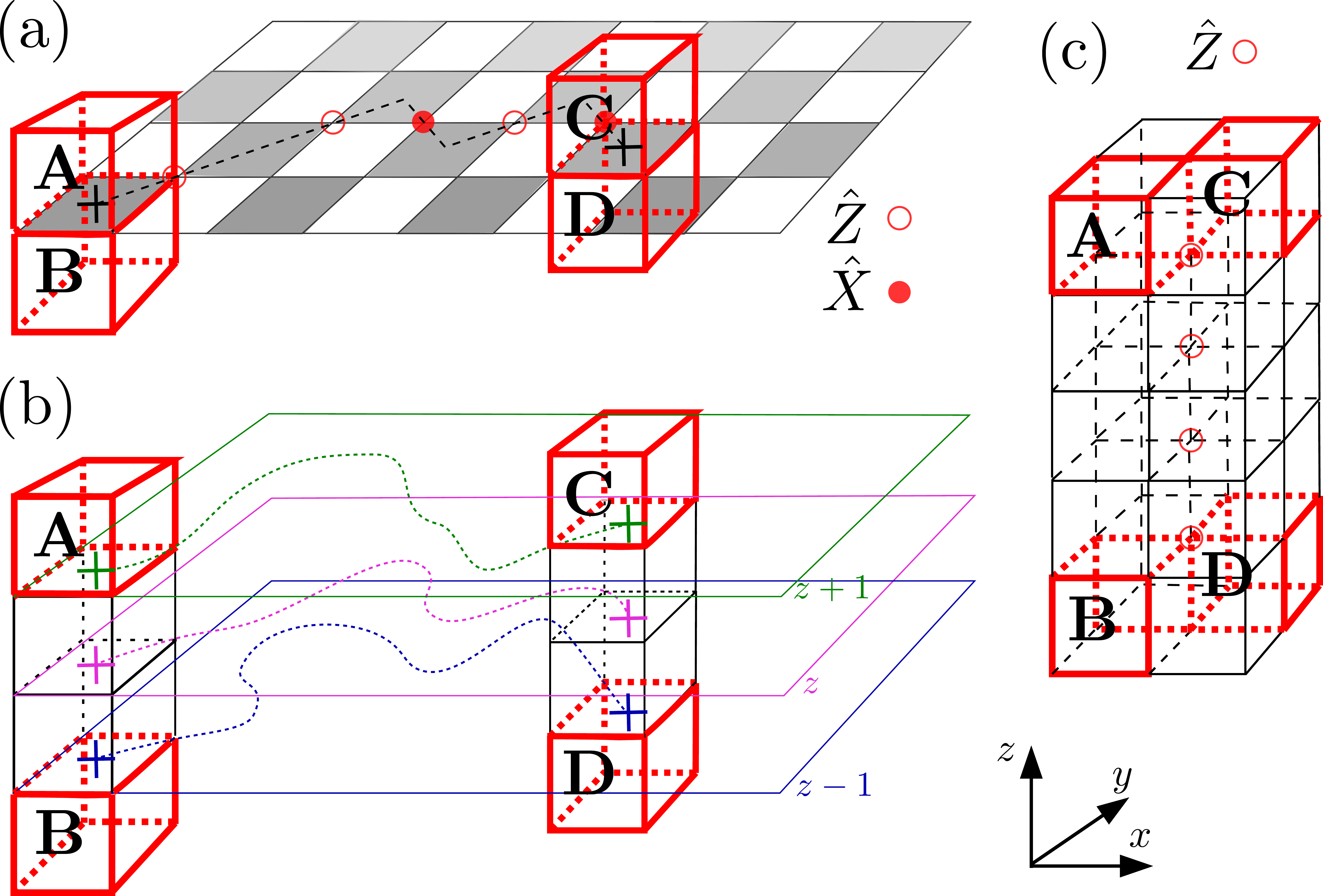}
\caption{(a) \emph{Dimension-2 particles} (AB and CD, pairs of cubes with $Q_c=-1$ sharing an $xy$ face) are located at the ends of an open string operator (dotted line), which is realized as a product of $\hat{X}$ and $\hat{Z}$ indicated by filled and empty circles respectively. (b) Pairs of cubes making up dimension-2 particles can be split into individual fractons (A, B, C, and D) by stacking open strings in adjacent $xy$ planes such that their endpoints are directly above one another. The open strings can be freely deformed apart from their endpoints, and are indicated by the dotted lines. (c) \emph{Dimension-1 particles} are pairs of cubes, diagonal from one another, at the same height $z$: AC and BD.}
\label{fig:3dSurface}
\end{figure}

Acting on a single spin in the ground state with $\hat{X}$ or $\hat{Z}$ results in flipping eigenvalues of four $\hat{Q}_c$ operators to $-1$, as illustrated in Fig.~\ref{fig:3dFlip}. The procedure for taking the four excited cubes apart is as follows. Applying a series of $\hat{X}$ or $\hat{Z}$ operators along a segment of spins in an $xy$ plane, as shown in Fig.~\ref{fig:3dSurface}(a), results in a string operator, connecting diagonal plaquettes. This string operator is very similar to that of the two-dimensional toric code, except that each end of the string hosts two excitations: cubes that share a face in the $xy$ plane. Because string operators connect diagonal plaquettes, such pairs of excitations come in two possible flavors, $e$ and $m$, corresponding to the two colors of the checkerboard pattern in the $xy$ plane. Additionally, we can separate a vertical pair of cubes by flipping the value of the bottom $Q_{xy}$ of the lower cube, or the top $Q_{xy}$ of the upper one. Such an operation brings the lower (upper) cube back to $Q_c=+1$, and shifts the $Q_c=-1$ excitation to its neighbor along $\hat{z}$. This newly-flipped $Q_{xy}(z\mp1)$ is located in the same column as another excited cube, and is located at an endpoint of an open string lying in the $xy$ plane at $z\mp1$. In the absence of open boundaries in the $\hat{x}$ or $\hat{y}$ directions, the string needs to have a second endpoint. The only position we can place it without increasing the quantity of $Q_c=-1$ excitations in the system is directly below (above) another excited cube, such that the second pair of vertical charges is separated by one layer of cubes in the $\hat{z}$ direction as well. We can continue this process, bringing the four excited cubes, termed \emph{fractons}, further apart as shown in Fig.~\ref{fig:3dSurface}(b). We note that the string operators defined in the $xy$ planes, such as the one shown in Fig.~\ref{fig:3dSurface}(a), commute with the Hamiltonian Eq.~(\ref{eq:H3d}) everywhere except for their endpoints (where they anti-commute). It follows that acting on a ground state of the Hamiltonian Eq.~(\ref{eq:H3d}) with such a string operator produces an excited eigenstate. However, the shape of a string operator can be freely deformed (while keeping its $e$ or $m$ flavor intact), and as long as the string's endpoints are fixed, the deformation results in the same eigenstate. This is the result of the special nature of the model's ground states. The latter can be thought of as superpositions of closed strings acting on a reference spin configuration, with phase prefactors that depend on the ground state's $Q_{xy}$ quantum numbers. The construction of these ground states proceeds in a manner similar to that of the toric code model discussed in the Appendix. For the fractons living at the corners of a two-dimensional region with a certain height (i.e., involving multiple $xy$ strings at adjacent layers), the string at each layer can be deformed individually. If a deformed string crosses a $Q_{xy}=-1$ line, it picks up a $-1$ phase factor relative to the original state. 

The four-cube single spin flip excitation has a particular flavor ($e$ or $m$), which is preserved when dimension-2 particles are taken apart since string operators in $xy$ planes only connect diagonal plaquettes. Vertical pairs of cubes can only be separated along $\hat{z}$. Therefore, all four fractons localized at the corners of the operator with two-dimensional support share the same flavor. It is impossible to alter the position of a single fracton at a time without paying an additional energy cost. Instead, fractons can only be shifted in pairs: either in two-dimensional $xy$ planes [Fig.~\ref{fig:3dSurface}(a)], or along straight lines in the $\hat{z}$ direction [Fig.~\ref{fig:3dSurface}(c)]. Adopting already existing terminology \cite{PhysRevB.92.235136}, we call such composite structures dimension-2 and dimension-1 particles respectively.

\begin{figure}
\includegraphics[width=0.8\columnwidth]{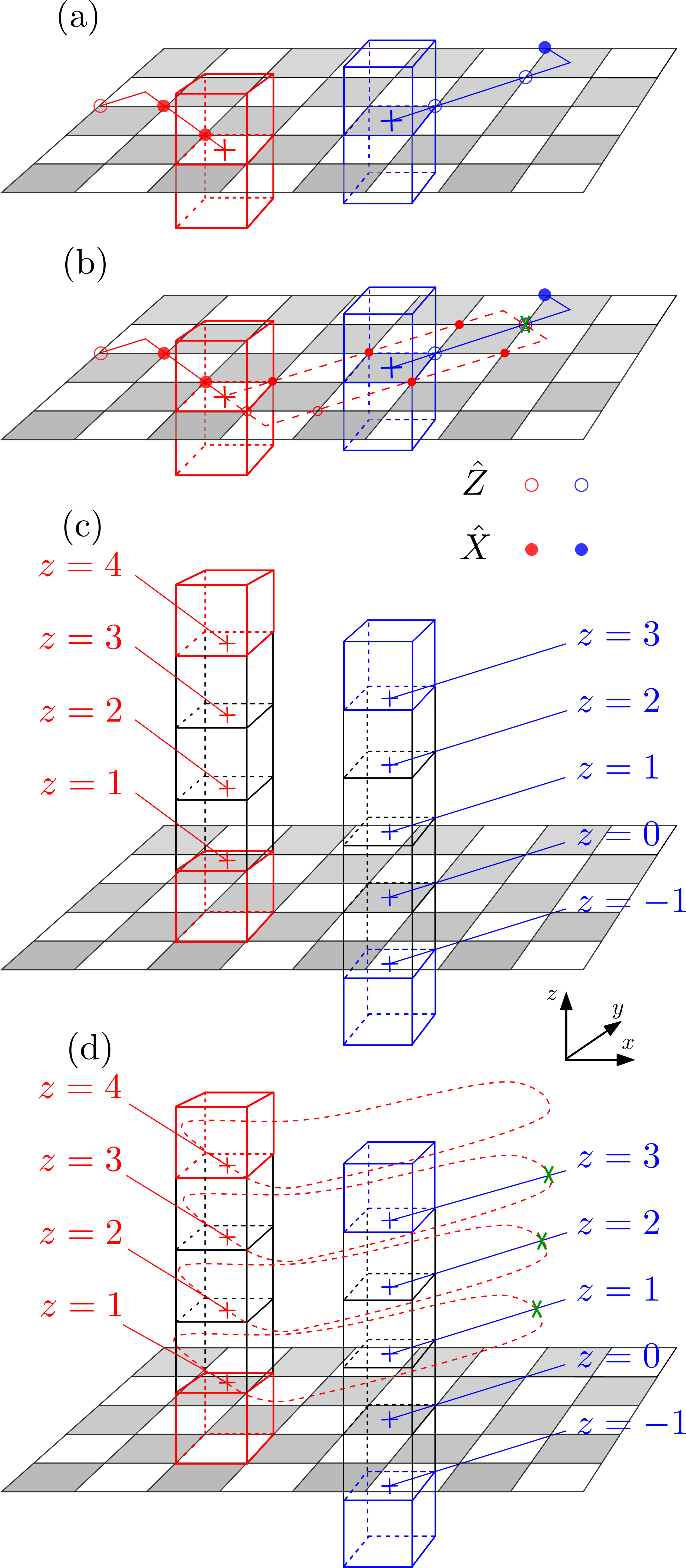}
\caption{(a) Two dimension-2 particles of different flavors, each consisting of a pair of adjacent excited cubes, are shown. The red ($m$ flavored) and blue ($e$ flavored) strings schematically indicate the presence of a second dimension-2 particle of the same flavor at the other end of the string. (b) The red dotted string indicates that the $m$ flavored dimension-2 particle has been moved around the $e$ particle in a closed loop. The green cross designates the crossing of the $e$ and $m$ strings. (c) Both $m$ (red) and $e$ (blue) dimension-2 particles take the form of rods extended along the $\hat{z}$ direction. (d) The four dotted red cycles stacked on top of one another along $\hat{z}$ depict moving of the $m$-flavored rod in space in a closed loop.This winding process leads to three $e$-$m$ string crossings, indicated by green crosses.}
\label{fig:winding}
\end{figure}

Particles propagating solely in one dimension cannot be exchanged with one another, therefore it does not make sense to speak about their exchange statistics. Since dimension-2 particles are constrained to move in two dimensions only, they may be exchanged and in principle can be anyons. Abelian anyons are detected via a non-trivial phase factor gained by the wavefunction upon a double exchange of two particles \cite{PhysRevLett.49.957}. This double exchange can be performed by winding one particle around the other. Consider an example shown in Fig.~\ref{fig:winding}(a-b). The initial state with $e$ and $m$ dimension-2 point-like particles is shown in Fig.~\ref{fig:winding}(a). The winding process consists of moving one of the particles, $m$, in space in a closed loop (returning to its initial position), such that the loop encloses the $e$ particle [Fig.~\ref{fig:winding}(b)]. Since closed strings are invisible, the resulting state looks identical, but has picked up a phase factor of $-1$ due to the $m$ loop and the $e$ string anti-commuting when they cross. Therefore, just as in the toric code model, $e$ and $m$ dimension-2 particles are mutual semions - anyons that acquire a minus sign upon braiding. However, one has to be careful when defining exchange statistics for dimension-2 particles that are extended along the $\hat{z}$ direction, as depicted in Fig.~\ref{fig:winding}(c). The two portrayed rods share an odd number of $xy$ planes in between their pairs of fractons (three, to be precise). When the $m$ rod is taken around the $e$ rod as shown in Fig.~\ref{fig:winding}(d), this results in three factors of $-1$ picked up from the anti-commutation relations between the $m$ and $e$ strings in the $xy$ planes. It is evident that the final accumulated phase factor depends on both the heights of the dimension-2 rods, and their relative positions along $\hat{z}$.

\subsection{Stability of the fracton order}

\begin{figure}
\includegraphics[width=\columnwidth]{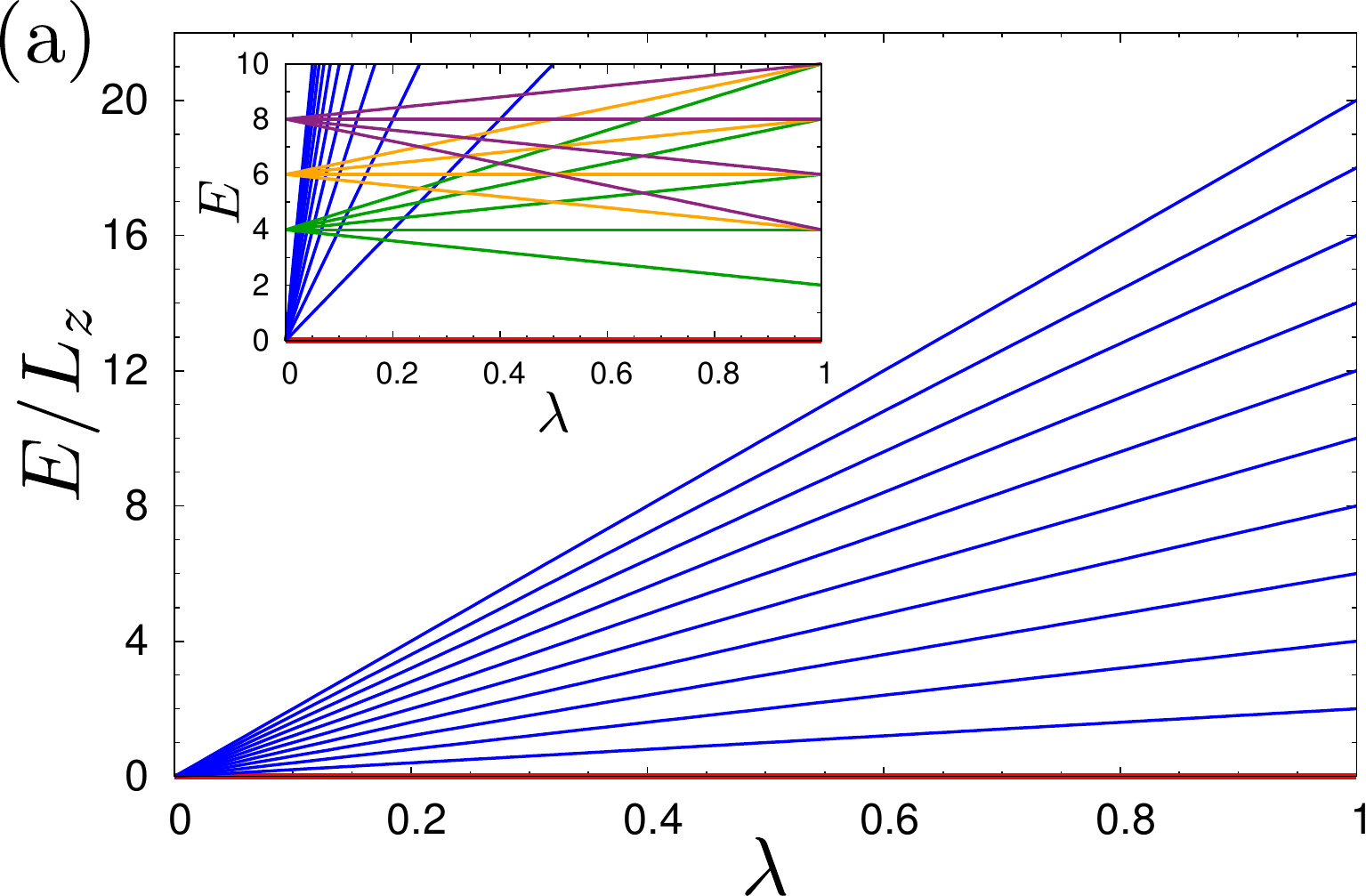}\\
\vspace{0.2cm}

\includegraphics[width=\columnwidth]{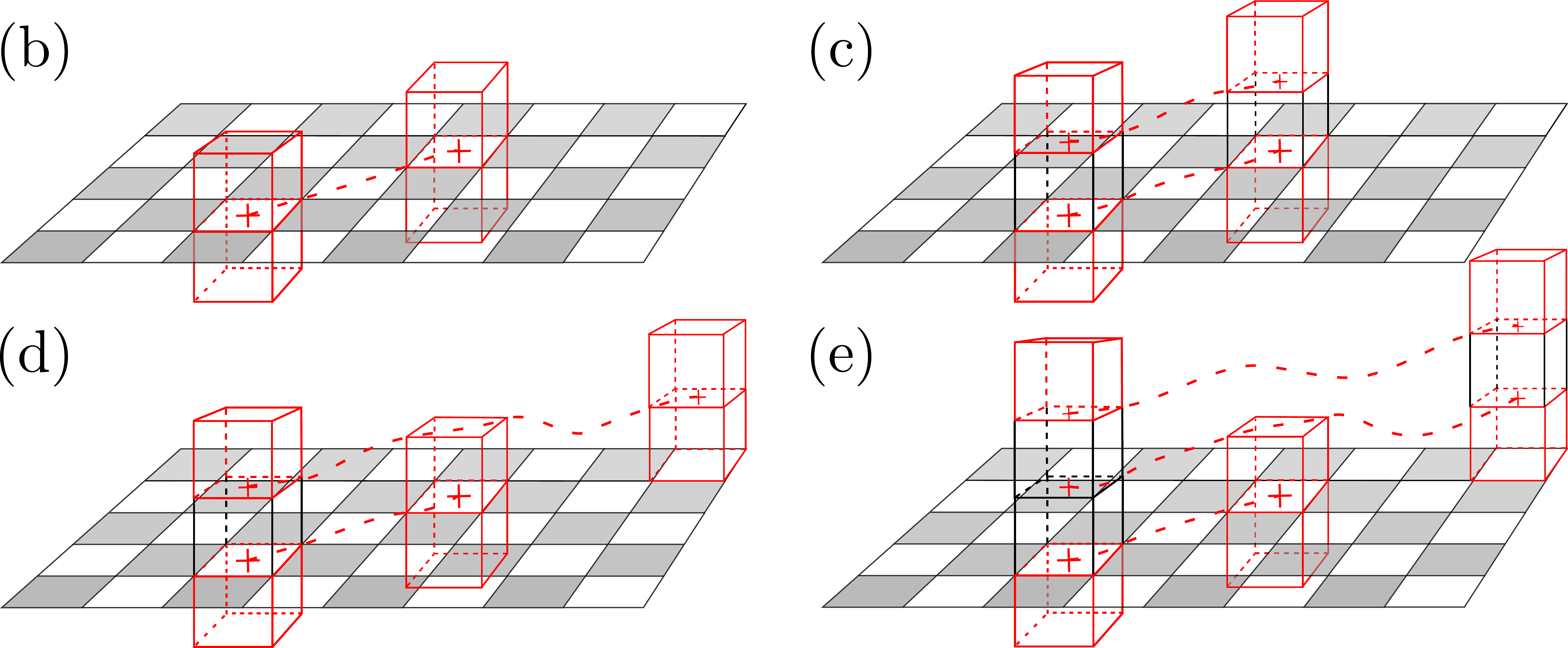}
\caption{{\it Upper panel} : Partial lifting of the ground state degeneracy of Eq.~(\ref{eq:Hdecoupled}) as a function of $\lambda$ for a system with $L_x\times L_y=20$. The energy level shown in red corresponds to the $2^{2L_z}$ ground states of the fracton model (\ref{eq:H3d}) with all $Q_{xy}$ quantum numbers equal to $+1$. The gap to the other states indicated in blue closes at the fracton point, $\lambda=0$. Inset shows how the first three excited energy levels of the fracton model are connected to the lowest five excited energy levels of the decoupled toric code planes.{ \it Lower panel} : Various examples of excited states. Excited cubes are in red and excited plaquettes have a red cross. Red crosses connected by a string lie in the same $xy$ plane. (b) and (c) have four excited cubes (corresponding to $E(\lambda=0)=4$) while having a different number of excited plaquettes corresponding respectively to $E(\lambda=1)=2$ and $E(\lambda=1)=4$. (d) and (e) have six excited cubes (corresponding to $E(\lambda=0)=6$) but have a different number of excited plaquettes corresponding respectively to $E(\lambda=1)=4$ and $E(\lambda=1)=6$.} 
\label{fig:gap}
\end{figure}

Our model exhibits both spontaneous symmetry breaking (along the $\hat{z}$ direction) and topological order \cite{1742-5468-2013-10-P10024}. For each symmetry-breaking configuration of $Q_{xy}$ quantum numbers of the ground states, there are $2^{2L_z}$ states that are locally indistinguishable and, therefore, topologically degenerate. Since the configuration of $Q_{xy}$ can be measured via local $\hat{Q}_{xy}$ operators, the same operators may be used to lift the non-topological part of the degeneracy. Assuming the topological order of our model to be stable (an assumption shown to be correct by the end of this Section), the presence of point-like excitations in the three-dimensional system suggests that if the fracton order gets destroyed, it is to be replaced with decoupled two-dimensional topologically ordered planes. To investigate this further, consider a Hamiltonian that allows to move between our fracton model Eq.~(\ref{eq:H3d}) and decoupled toric code planes:
\begin{equation}
H(\lambda)=-\frac{(1-\lambda)}{2}\sum_{c}\hat{Q}_c-\frac{\lambda}{2}\sum_{z}\sum_{p}\hat{Q}_{xy}(z)
\label{eq:Hdecoupled}
\end{equation}
where $\lambda$ is a parameter ranging from 0 to 1, and the second term is the sum of two-dimensional toric code Hamiltonians over the $xy$ planes of the cubic lattice (index $p$ corresponds to square plaquettes located in the $xy$ planes). 
When $\lambda\neq0$, the toric code part of Eq.~(\ref{eq:Hdecoupled}) partially lifts the ground state degeneracy of the fracton model Eq.~(\ref{eq:H3d}) by choosing the $2^{2L_z}$ states that have all $Q_{xy}(z)=+1$. The other states acquire energy costs that grow linearly with $\lambda$, as shown in Fig.~\ref{fig:gap}. Since the gap remains open for $\lambda\neq0$, the model in Eq.~(\ref{eq:Hdecoupled}) supports two regimes: the decoupled toric code planes' phase ($\lambda>0$) and the fracton point at $\lambda=0$. Since we know the toric code model to be stable to local perturbations \cite{toricKitaev}, we conclude that, while the topological order of our model is stable, its three-dimensional fracton nature is a fine-tuned point.

The inset in Fig.~\ref{fig:gap} shows the crossings of the energy levels as Eq.~(\ref{eq:Hdecoupled}) varies from the fracton point ($\lambda=0$) to the limit of fully decoupled toric code planes ($\lambda=1$). We note that the energy spectrum at $\lambda=1$ is equidistant in steps of 2 from 0 to $L_x\times L_y\times L_z$, whereas the spectrum at $\lambda=0$ is similar yet missing the lowest excited level equal to 2.

\section{Open boundary conditions}
\label{sec:open}

\subsection{Open boundaries in the $\hat{z}$ direction}

\begin{figure}[h!]
\includegraphics[width=0.8\columnwidth]{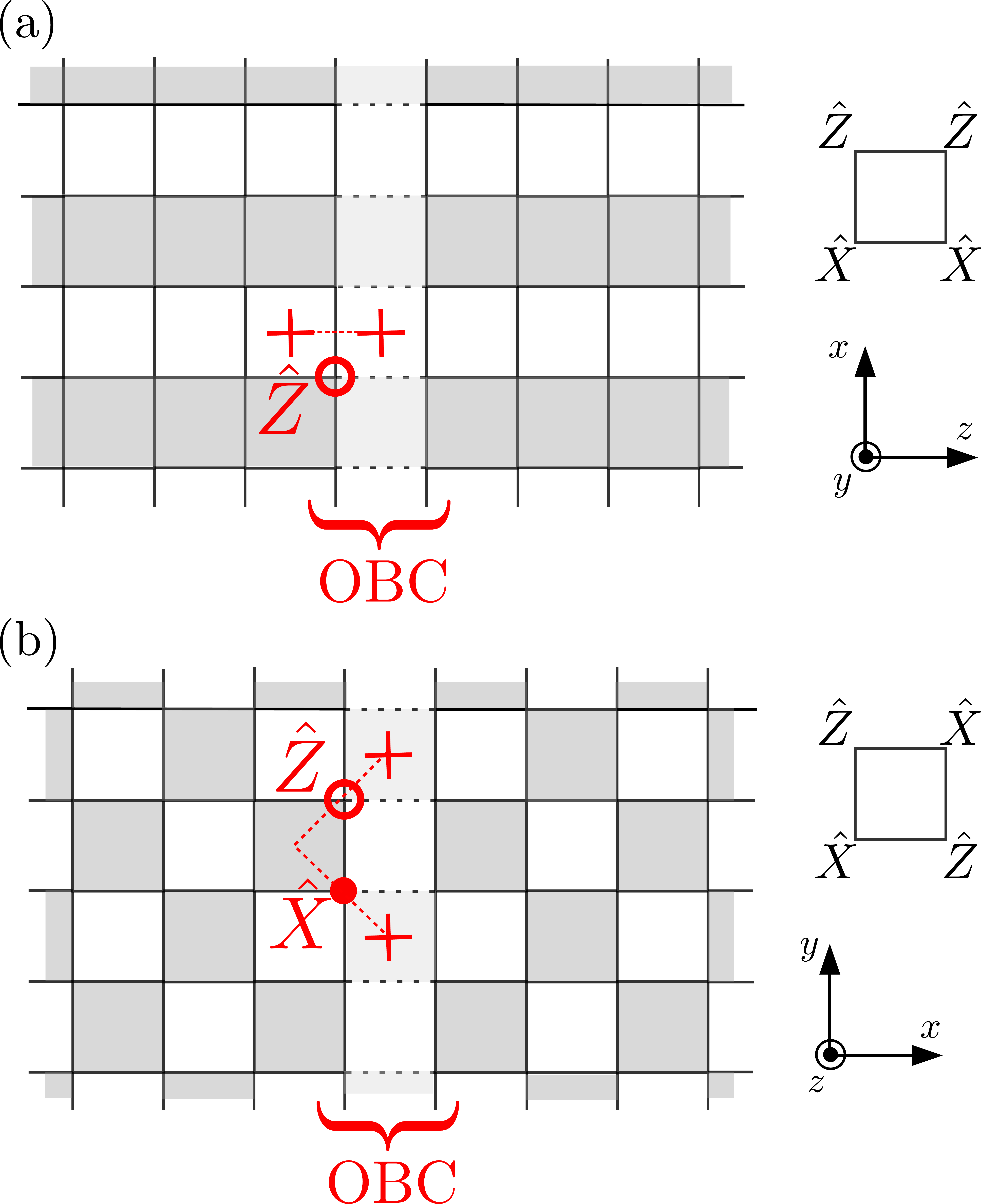}
\caption{Gray and white regions denote $e$ and $m$ flavors of the cubes projected onto 2D planes. The initial states having $Q_c=+1$ everywhere, cubes with $Q_c=-1$ are denoted by red crosses. Imposing an open boundary in a certain direction amounts to removing a plane of cubes perpendicular to that direction from the Hamiltonian (\ref{eq:H3d}). (a) It is impossible to change the eigenvalues of $\hat{Q}_c$ in the ``missing'' boundary layer perpendicular to $\hat{z}$ without introducing additional charges elsewhere in the bulk of the system. (b) Values of $\hat{Q}_c$ in the boundary layers perpendicular to $\hat{x}$ and $\hat{y}$ can be changed by acting on the system with open strings with both endpoints located at the surface.}
\label{fig:3dBoundary}
\end{figure}

So far our discussion has been limited to the case of periodic boundary conditions. Consider imposing open boundary conditions in the $\hat{z}$ direction. We may view such a system as one with periodic boundary conditions that has one layer of cubes, parallel to the $xy$ plane, absent from the Hamiltonian (\ref{eq:H3d}). One of the consequences of this change is that we are now able to ``expel'' pairs of fractons from the system by bringing them out to either of its surfaces. This has a direct effect on the excited states' spectrum of our model: it acquires an additional energy level with two $Q_c=-1$ cubes. 
The ground state degeneracy, however, remains unaffected, as introducing zero-energy $Q_c=-1$ charges into the ``missing" layer of cubes requires placing additional, energetically costly, charges into the bulk of the system [Fig.~\ref{fig:3dBoundary}(a)]. This conclusion is supported by counting the number of encoded qubits following the earlier prescription:  the number of spins, $L_x\times L_y\times L_z$, remains unchanged with the imposition of open boundary conditions, whereas the number of cubes (and, therefore, $Q_c=+1$ ground state constraints) is reduced to $L_x\times L_y\times( L_z-1)$. Since vertical pairs of fractons no longer have to come in pairs, only the dependencies between $\hat{Q}_c$ in the layers remain:
\begin{equation}
\prod_{c\in e}Q_c=+1 \mathrm{~and~} \prod_{c\in m}Q_c=+1
\end{equation}
per each of the $L_z-1$ layers of cubes. The ground state degeneracy therefore remains $\mathcal{D}=2^{L_x\times L_y +2(L_z-1)}$. Additionally, the surface degeneracy is still non-local, as going from one ground state to another requires flipping $L_z$ spins.

\subsection{Open boundaries in the $\hat{x}$ or $\hat{y}$ directions}

The situation changes drastically once open boundary conditions are imposed in the $\hat{x}$ and/or $\hat{y}$ directions. Again, we may think of the boundary as a layer of cubes, parallel to $yz$ or $xz$, that does not appear in the Hamiltonian (\ref{eq:H3d}). We can introduce $Q_c=-1$ charges in this plane without any energy cost, moreover, we may do so locally, flipping as few as two adjacent spins located on the surface, as shown in Fig.~\ref{fig:3dBoundary}(b). Such local spin flips, that generate zero energy modes, are simply $xy$ strings starting and terminating right outside of the system's boundary. It follows that for a system with boundary conditions that are open in $\hat{x}$ and periodic elsewhere, the number of surface zero modes is $2^{L_y\times L_z}$. Indeed, we arrive at the same result by calculating the total ground state degeneracy from the difference in the number of spins and independent constraints on the model's ground state. In this example, number of spins is $L_x\times L_y\times L_z$, number of cubes is $(L_x-1)\times L_y\times L_z$. In the presence of an open surface parallel to $yz$, there are only the column relations of the form (\ref{eq:crelation}), and there are  $(L_x-1)\times L_y$ of them. This results in the following expression for the ground state degeneracy: $\mathcal{D}=2^{(L_y\times L_z+(L_x-1)\times L_y)}$, where $2^{(L_y\times L_z)}$ arises due to the surface zero energy modes.

\section{Conclusion}

In this article, we start with a layered system made up of two-dimensional toric code planes \cite{toricWen,toricKitaev}, and couple the plaquette terms on adjacent planes so that we arrive at a point that cannot be connected to the toric code phase without opening up a gap in the energy spectrum. We derive the intuition for the resulting model defined on the cubic lattice from the two-dimensional physics of the toric code. We show that our three-dimensional model exhibits fracton topological order. Its excitations fractionalize into immobile pointlike particles, fractons, pairs of which can be combined into composite excitations that move either  in a straight line along the $\hat{z}$ direction, or freely in the $xy$ plane at a set height $z$. The ability to combine fractons into mobile particles that move in spaces with reduced dimensionality is common across many fracton models. We find that the presence of zero energy modes on the surfaces perpendicular to $\hat{x}$ and $\hat{y}$ directions, and the lack of them on the surfaces normal to $\hat{z}$, can be explained using the properties of the composite mobile excitations made up of pairs of fractons. 

One promising direction for future work involves investigating the effects of introducing defects into our model, including the so-called \emph{twists} -- defects related to a symmetry present in the anyon model, that give rise to more complex types of anyon excitations \cite{PhysRevLett.105.030403}. 

The authors thank Yuan Wan, Maurizio Fagotti, Oleg Tchernyschyov, Kirill Shtengel, Tim Hsieh, and Gabor Halasz for useful discussions. O.~P. gratefully acknowledges the hospitality of Perimeter Institute for Theoretical Physics, where this work was initiated. O.~P. was supported by LabEX ENS-ICFP: ANR-10-LABX-0010/ANR-10-IDEX-0001-02 PSL*. N.~R. was supported by Grant No.~ANR-17-CE30-0013-01.

\bibliography{toricbib}

\appendix 

\section{Brief review of the 2D toric code}

\begin{figure}
\includegraphics[width=\columnwidth]{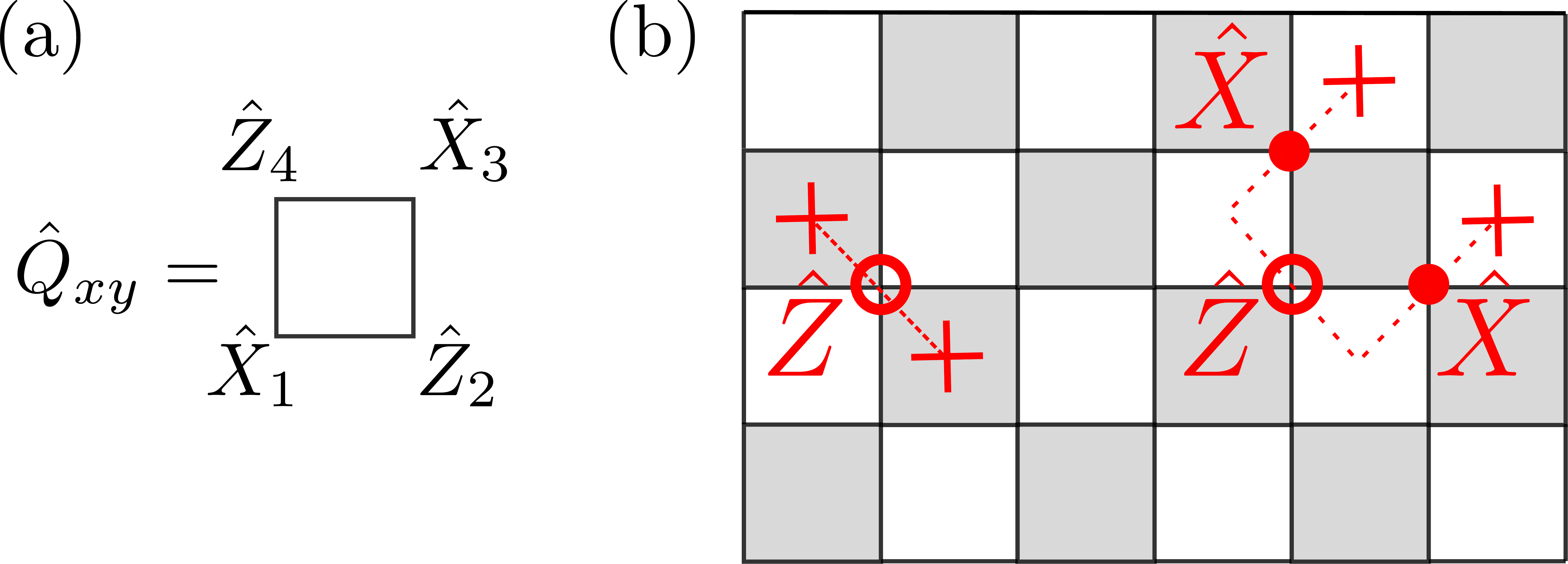}
\caption{The toric code (equivalently, Wen's plaquette model) in two dimensions. (a) Building blocks of the Hamiltonian (\ref{eq:H2d}) are plaquette operators $\hat{Q}_{xy}$. (b) Excitations (indicated by red crosses) are located at the ends of string operators consisting of successive $\hat{X}$ and $\hat{Z}$, shown as dotted lines connecting the centers of diagonal plaquettes.}
\label{fig:toric2d}
\end{figure}

We consider Wen's plaquette model \cite{toricWen} defined for spins $1/2$ living on the sites of a square lattice. This model can be cast into the form of Kitaev's toric code \cite{toricKitaev} simply by moving spins from the square lattice sites to the edges of a new square lattice and performing unitary rotations on the Pauli matrices. Spins interact via four-spin plaquette operators $\hat{Q}_{xy}=\hat{X}_1 \hat{Z}_2 \hat{X}_3 \hat{Z}_4$, shown in Fig.~\ref{fig:toric2d}(a). The Hamiltonian is simply the sum of such terms
\begin{equation}
H=-\sum_{p}\hat{Q}_{xy},
\label{eq:H2d}
\end{equation}
taken over all square plaquettes $p$. All $\hat{Q}_{xy}$ commute with one another. Their two possible eigenvalues are $Q_{xy}=\pm1$. The ground state of the Hamiltonian (\ref{eq:H2d}) has all $Q_{xy}=+1$. Pairs of $Q_{xy}=-1$ excitations can be created via the application of Pauli operators $\hat{X}$ or $\hat{Z}$: each such operator flips the signs of the eigenvalues of $\hat{Q}_{xy}$ on diagonal plaquettes, as shown in Fig.~\ref{fig:toric2d}(b). Successive applications of $\hat{X}$ or $\hat{Z}$ bring the two excitations further apart. The corresponding operator, indicated by the dotted red string in Fig.~\ref{fig:toric2d}(b), that connects the centers of diagonal plaquettes whose $Q_{xy}$ have been flipped, comes in two flavors. If we color the square lattice plaquettes in a checkerboard pattern, the two flavors, $e$ and $m$, correspond to the dark and light shades of the squares in Fig.~\ref{fig:toric2d}(b) respectively. It follows that the $Q_{xy}=-1$ excitations also have either $e$ or $m$ flavor, and are located at the ends of the corresponding string operators.

\begin{figure}
\includegraphics[width=\columnwidth]{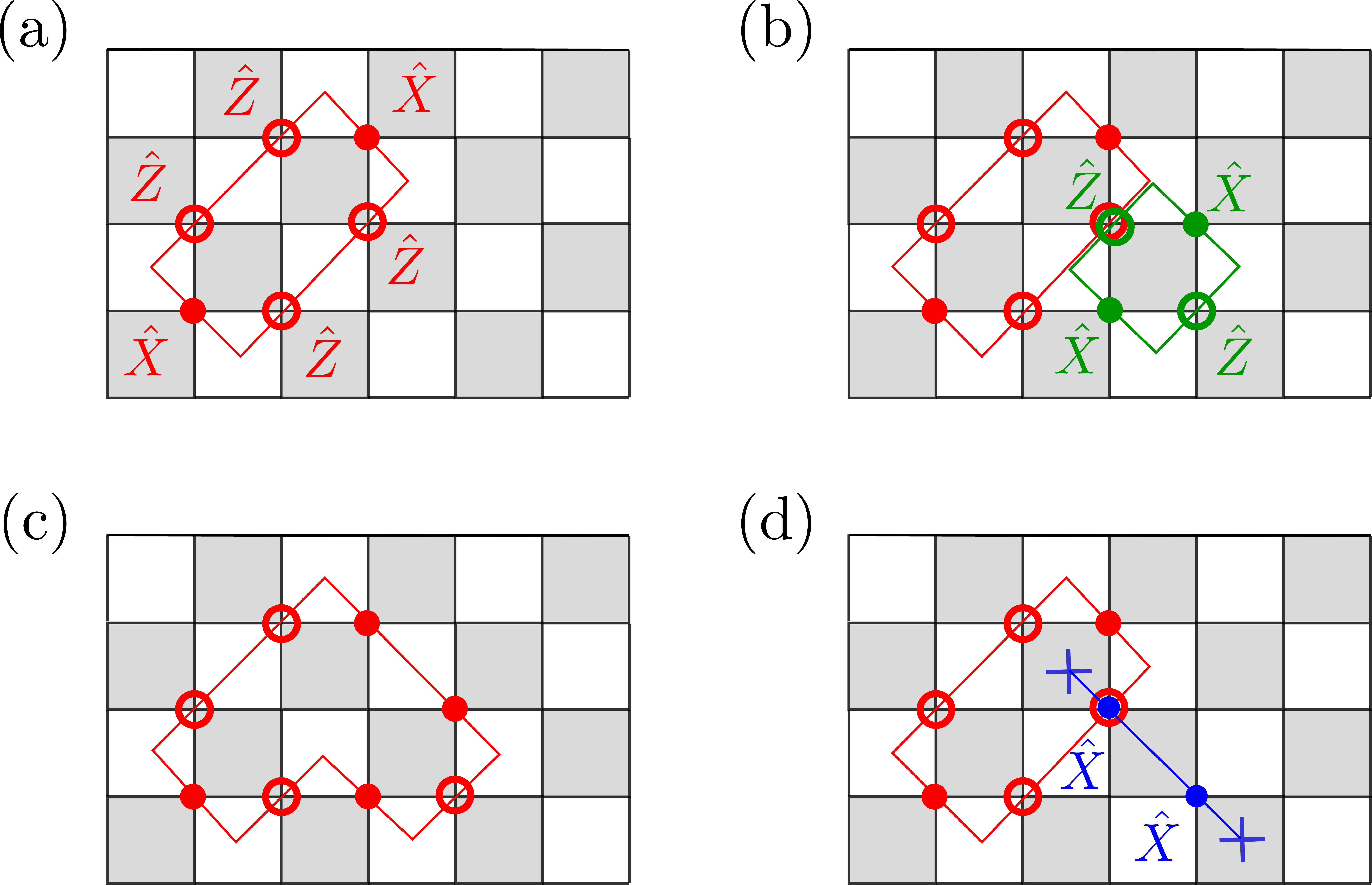}
\caption{(a) A closed string of $e$ flavor (red). (b) An adjacent single-plaquette closed string of $m$ flavor (green), equal to the corresponding $\hat{Q}_{xy}$ operator. (c) The product of (a) and (b) is also a closed string of $e$ flavor. (d) A closed $e$ string enclosing an $m$-flavored excitation.}
\label{fig:soup}
\end{figure}

Since excitations live at the strings' endpoints, applying a closed string operator does not result in any energy increase. It can be shown that closed strings commute with the Hamiltonian (\ref{eq:H2d}). In fact, the closed string picture can give us important insights into the nature of the toric code ground states. Consider a closed string shown in Fig.~\ref{fig:soup}(a) that is applied to some reference spin configuration. Let us look at what happens when we act on the resulting state with a $\hat{Q}_{xy}$ operator associated with an adjacent plaquette, indicated in green in Fig.~\ref{fig:soup}(b) (note that $\hat{Q}_{xy}$ itself can be viewed as the minimal closed string). The red closed string and the green plaquette result in a closed string with a different shape, as pictured in Fig.~\ref{fig:soup}(c). Since applying $\hat{Q}_{xy}$ to closed strings results in their deformation, it follows that a superposition of all closed strings connected by $\hat{Q}_{xy}$ constitutes a ground state of the Hamiltonian (\ref{eq:H2d}). Since all $\hat{Q}_{xy}$ operators commute among each other, applying them to excited states containing open strings does not change the locations of the $Q_{xy}=-1$ excitations. It does, however, change the shape of the string apart from its endpoints. Since we are applying a well-defined string operator (e.g., the one shown in Fig.~\ref{fig:toric2d}(b) in red) to a superposition of closed strings, the string itself is invisible: any choice of its shape results in the same physical state as long as the endpoints are held fixed.

\begin{figure}
\includegraphics[width=\columnwidth]{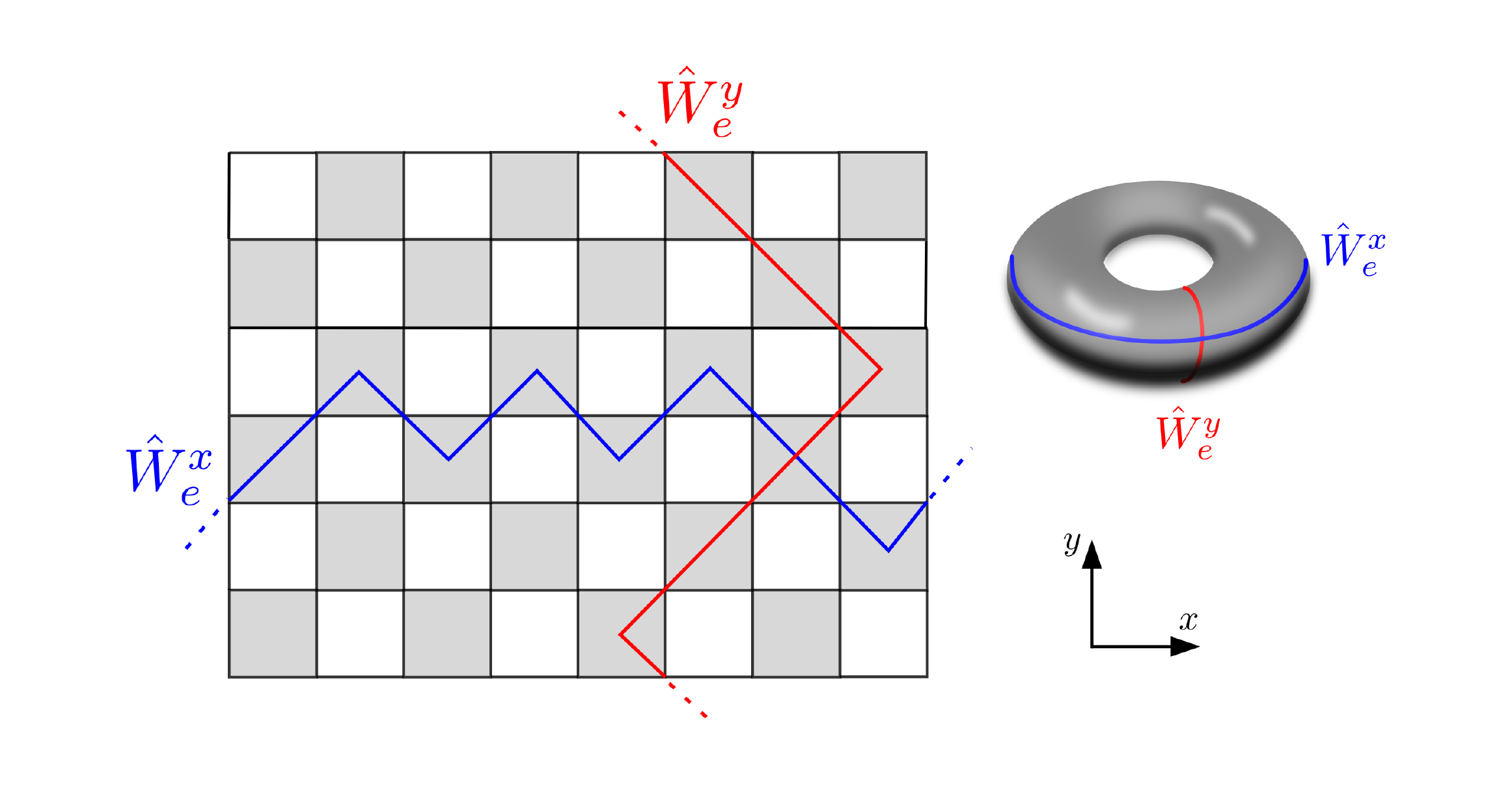}
\caption{Two nontrivial closed strings in the toric code, whose eigenvalues can be used to specify one of the 4 possible ground states.}
\label{fig:toric2dGS}
\end{figure}

It can be shown that, for an even$\times$even system embedded on a torus, there exist four ground states that are not connected by the local $\hat{Q}_{xy}$ operators. These states can be identified by the eigenvalues $W^{x}_{e}$ and $W^{y}_{e}$ of the following two non-local operators: $\hat{W}^{x}_{e}$ and $\hat{W}^{y}_{e}$, closed strings of the same flavor that wrap around the two non-contractible loops of the torus shown in Fig.~\ref{fig:toric2dGS}. One can only switch between these four states via other non-local operators, such as closed non-contractible strings of a different flavor.

The presence of topological degeneracy in the ground state is a smoking gun for the existence of excitations with non-trivial exchange statistics. Indeed, while it can be shown that both $e$ and $m$ plaquette excitations are bosons with respect to themselves, they are actually mutual anyons with respect to each other. A double exchange of two Abelian anyons should give rise to an overall phase factor. A closed string of one flavor, enclosing an open end of a string with the other flavor, is shown in Fig.~\ref{fig:soup}(d). Strings of different flavors anti-commute when they cross, therefore the closed string in question measures the parity of the enclosed charge. Since a double exchange of two particles is topologically equivalent to having one particle go around the other in a closed loop, it follows that a double exchange of an $e$ and an $m$ particle picks up a factor of $-1$, i.e. the two are mutual anyons.

\end{document}